\documentclass[12pt]{article}

\usepackage{epsfig}

\def\be{\begin{equation}}
\def\ee{\end{equation}}
\def\mueg{\mu^- \to e^- \gamma}
\def\taueg{\tau^- \to e^- \gamma}
\def\taumug{\tau^- \to \mu^- \gamma}
\newcommand {\eq} [1] {Eq.~(\ref{#1})}
\newcommand {\eqz} [2] {Eqs.~(\ref{#1}) and (\ref{#2})}
\newcommand {\fig} [1] {Fig.~\ref{#1}}
\newcommand {\misset}  {E_T\hspace{-4mm}/ \hspace{1mm}}

\begin{document}

\begin{flushright}
\small HEPHY-PUB-764/02 \\
ZU-TH 18/02
\end{flushright}
\begin{center}
{\bf \Large Large lepton flavour violating signals \\[0.1cm] in supersymmetric
       particle decays } \\[0.4cm]
{\large \underline{W.~Porod}$^{a}$ and W.~Majerotto$^{b}$
}\\[0.5cm] 
{\small
$^a$ Inst.~f\"ur Theor. Physik, Universit\"at Z\"urich, CH-8057 Z\"urich,
      Switzerland \\ \small
$^b$ Inst.~f.~Hochenergiephysik, \"Oster.~Akademie d.~Wissenschaften,
      A-1050 Vienna, Austria
}
\end{center}

\begin{abstract}
  We study lepton flavour violating signals at a future $e^+ e^-$
  linear collider within the general MSSM, allowing for the most
  general flavour structure. We demonstrate that there is a large
  region in parameter space with large signals, while being
  consistent with present experimental bounds on rare lepton decays
  such as $\mueg$.  In our analysis, we include all
  possible signals from charged slepton, sneutrino,
  neutralino, and chargino production and decay. 
  We also consider the background 
  from the Standard Model {\it and} the MSSM. We find that in general
  the signature $e \tau \misset$ is the most pronounced one. 
\end{abstract}

\section{Introduction}

Neutrino experiments have established the existence of lepton flavour
violation (LFV). On the one side, the results of Super-Kamiokande 
yield an almost maximal
mixing between $\nu_\mu$ and $\nu_\tau$ \cite{Nu1}
and also the latest results of SNO \cite{Nu2}
suggest that also the $\nu_e-\nu_\mu$ sector contains a large
mixing, whereas the third mixing between $\nu_e$ and $\nu_\tau$
 has to be small \cite{Nu3}.
On the other side,
there are stringent constraints on LFV in the charged
lepton sector, the strongest being 
$BR(\mueg) < 1.2 \times 10^{-11}$ \cite{Groom:2000in}. Others are
$BR(\mu^- \to e^- e^+ e^-) < 10^{-12}$, $BR(\taueg) < 2.7 \times 10^{-6}$,
$BR(\taumug) < 1.1 \times 10^{-6}$.
The Standard Model can account for the lepton flavour 
conservation in the charged lepton sector, but has to be extended to
account for neutrino masses and mixings, e.g.~by the see-saw mechanism
and by introducing heavy right-handed Majorana neutrinos
\cite{seesaw}.

In general, a gauge and supersymmetric invariant theory does neither conserve
total lepton number $L=L_e + L_\mu + L_\tau$ nor individual lepton number
$L_e$,  $L_\mu$ or  $L_\tau$. One usually invokes R-parity symmetry, which
forces total lepton number conservation but still allows the violation of
individual lepton number, e.g.~due to loop effects in $\mueg$ \cite{ref11}.
The Minimal Supersymmetric Standard Model (MSSM) with R-parity
conservation embedded in a GUT theory induces LFV
\cite{ref6} at the weak scale. This is a consequence of having
leptons and quarks in the same GUT multiplet and of the quark flavour
mixing due to the CKM matrix.  A general analysis of flavour changing
neutral (FCNC) effects in K- and B-meson as well as in lepton physics
was recently performed in \cite{ref7}.

Moreover, in the MSSM a large $\nu_\mu$-$\nu_\tau$ mixing can lead to a 
large  $\tilde \nu_\mu$-$\tilde \nu_\tau$ mixing
via renormalisation group equations \cite{ref10}. This leads to clear LFV
signals in slepton and sneutrino production
and in the decays of neutralinos and charginos into sleptons and sneutrinos
at the LHC \cite{ref12} and at future lepton colliders
\cite{ref8}. Signatures due to $\tilde e_R$-$\tilde \mu_R$ mixing
were discussed in \cite{ref14}. In all these studies, it has been assumed that
only one lepton flavour violating term dominates.

In this contribution, we present the results of \cite{ego} 
where we have studied the consequences of LFV
in the sfermion sector at future $e^+ e^-$ colliders, and we give additional 
new
results.   Assuming the {\it most general} mass matrices for
sleptons and sneutrinos, we demonstrate that
large signals are expected while at the same time respecting present 
bounds on rare lepton decays.

\section{Lepton Flavour Violation in the MSSM}

The most general charged slepton $6\times 6$
mass matrix including left-right mixing
as well as flavour mixing is given by:
\begin{equation}
  M^2_{\tilde l} = \left(
    \begin{array}{cc}
      M^2_{L,ij} + \frac{1 }{2} v^2_d Y^{E*}_{ki} Y^{E}_{kj}
       + D_L  \delta_{ij}  &
        \frac{1}{\sqrt{2}} (v_d A_{ji} - \mu^* v_u Y^{E*}_{ij} )   \\
     \frac{1 }{\sqrt{2}} (v_d A^*_{ji} - \mu v_u Y^{E}_{ij}) & 
        M^2_{R,ij} + \frac{1}{2} v^2_d Y^{E}_{ik} Y^{E*}_{jk} 
      -  D_R \delta_{ij}   \\
     \end{array} \right) \, 
  \label{eq:sleptonmass}
\end{equation}
with $D_L =\frac{1}{8}\left( {g'}^2 -  g^2 \right) (v^2_d - v^2_u)$ and
$D_R=\frac{1}{4} {g'}^2  (v^2_d - v^2_u)$.
The indices $i,j,k=1,2,3$ characterize the flavors $e,\mu,\tau$.
$M^2_{LL}$ and $M^2_{RR}$ are the soft SUSY breaking mass matrices for
left and right sleptons, respectively. $A_{ij}$ are the trilinear soft
SUSY breaking couplings of the sleptons and Higgs bosons.
The physical mass eigenstates states $\tilde l_n$ are given by 
$\tilde l_n = R^{\tilde l}_{nm} \tilde l'_m$ with 
$l'_m = (\tilde e_L, \tilde \mu_L, \tilde \tau_L,
          \tilde e_R, \tilde \mu_R, \tilde \tau_R)$.
Similarly, one finds for the sneutrinos
\begin{equation}
  M^2_{\tilde \nu,ij} =  M^2_{L,ij} \textstyle
  + \frac{1}{8} \left( g^2 + {g'}^2 \right) (v^2_d - v^2_u) \delta_{ij}
  \label{eq:sneutrinomass}
\end{equation}
with the physical mass eigenstates 
$\tilde \nu_i = R^{\tilde \nu}_{ij}\tilde \nu_j'$ and 
$\tilde \nu_j' = (\tilde \nu_e, \tilde \nu_\mu, \tilde \nu_\tau) $.
The relevant interactions for this study are given by:
\begin{eqnarray}
  \label{eq:CoupChiSfermion}
 {\cal L} &=& \bar l_i ( c^L_{ikm} P_L + c^R_{ikm} P_R)
               \tilde \chi^0_k \tilde l_m  
    +  \bar{l_i} (d^L_{ilr} P_L + d^R_{ijr} P_R)
           \tilde \chi^-_l \tilde{\nu}_r
 + \bar{\nu_i} ( e^R_{ilm} P_R) \tilde \chi^+_l \tilde{l}_m 
+ h.c.
\end{eqnarray}
with
\begin{eqnarray}
 c^L_{ikm} &=&
  - \sqrt{2} g' ( R^{\tilde l}_{m,i+3} )^*  N_{k1}^*
  - ( R^{\tilde l}_{mi} )^*  Y^E_{ii} N_{k3}^* \\
 c^R_{ikm} &=&
  ( R^{\tilde l}_{ki} )^* \frac{g'  N_{k1} + g N_{k2} }{\sqrt{2}}
   - ( R^{\tilde l}_{m,i+3} )^*  Y^E_{ii} N_{j3} \\ 
 d^L_{ilr} &=& Y^E_{ii} ( R^{\tilde \nu}_{ri} )^* U^*_{l2} \, , \hspace{1cm}
 d^R_{ilr} = - g * ( R^{\tilde \nu}_{ri} )^*  V_{l1} \\ 
 e^R_{ilm} &=& \sum_r \left( - g ( R^{\tilde l}_{mr} )^* R^\nu_{ir} U_{l1}
         + Y^E_{rr} ( R^{\tilde l}_{m,r+3} )^* R^\nu_{ir} U^*_{l2} \right)
\end{eqnarray}
where we have chosen the basis where the charged lepton Yukawa coupling
is diagonal $Y^E_{ij} = \sqrt{2} m_i / v_d \delta_{ij}$. $R^\nu_{ij}$ is
the neutrino mixing matrix, $N$ diagonalises the neutralino mass matrix 
in the basis  $\tilde B, \tilde W_3, \tilde H^0_u, \tilde H^0_d$ and
$U$ and $V$ are the mixing matrices of the charginos.
The first two terms in \eq{eq:CoupChiSfermion}
give rise to LFV
signals whereas the last one will give rise to the SUSY background.

As mentioned above, most studies so far consider the case where only one
of the flavour mixing entries in Eqs.~(\ref{eq:sleptonmass}) and/or 
(\ref{eq:sneutrinomass}) is non-zero. It is the purpose of this study to allow
for all possible flavour violating entries  in 
\eqz{eq:sleptonmass}{eq:sneutrinomass} which are compatible
with the present bounds on lepton number violating processes, such as
$\mueg, e^- e^+ e^-$, $\taueg$, $\taumug$ and 
$Z\to e \mu, e \tau, \mu \tau$.
For definiteness, we have taken the first of the mSUGRA points of Snowmass' 01 
\cite{Georg}:  $M_{1/2} = 250$~GeV,
$M_0=100$~GeV, $A'_0=-100$~GeV, $\tan \beta = 10$ and sign$(\mu)=+$.
Note that $A'_0$ has  to be multiplied by the Yukawa couplings to get
the $A_{ij}$ parameters of \eq{eq:sleptonmass}. This leads to the
following slepton mass parameters at the electroweak scale: 
$M_{R_{11}}$ = 138.7~GeV,
$M_{R_{33}}$ = 136.3~GeV, $M_{L_{11}}$ = 202.3~GeV, $M_{L_{33}}$ = 201.5~GeV
and $A_{33}/Y^E_{33}$  = -257.3~GeV. Some typical masses are: 
$m_{\tilde e_R} = 146.9$~GeV, $m_{\tilde e_L} = 214.7$~GeV,
$m_{\tilde \nu_e} = 199.4$~GeV,
$m_{\tilde \tau_1} = 138.6$~GeV, $m_{\tilde \tau_2} = 217.7$~GeV,
$m_{\tilde \chi^+_1} = 193.6$~GeV, $m_{\tilde \chi^0_1} = 103.1$~GeV,
$m_{\tilde \chi^0_2} = 194.6$~GeV, $m_{A^0} = 395$~GeV and
$m_{\tilde t_1} = 407$~GeV
(the remaining masses are given in~\cite{ego}).
We keep all parameters fixed
except for the slepton parameters $M^2_L$, $M^2_R$ and $A$ where all
entries are varied in the whole range compatible with 
the experimental constraints.

\begin{figure}[t]
\setlength{\unitlength}{1mm}
\begin{picture}(170,58)
\put(0,-25){\mbox{\epsfig{figure=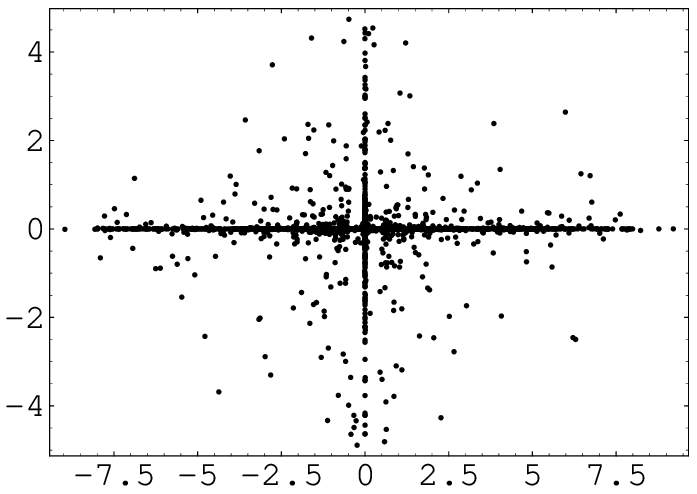,height=10.cm,width=7cm}}}
\put(85,-25){\mbox{\epsfig{figure=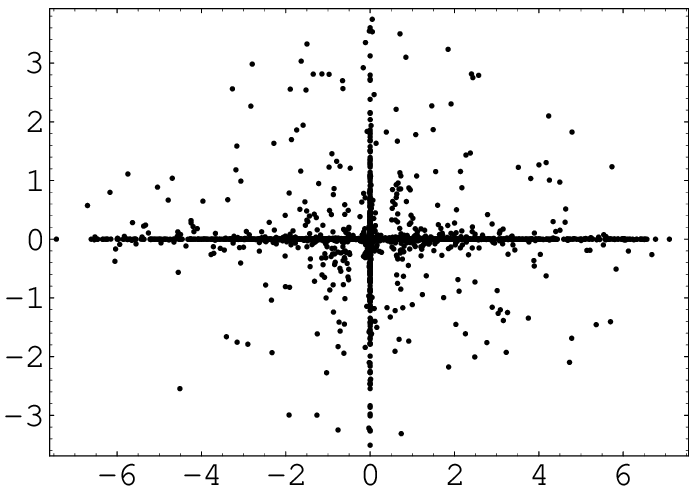,height=10.cm,width=7cm}}}
\put(-1,54){\mbox{\bf (a)}}
\put(5,53){\mbox{$M^2_{L,13} \cdot 10^3$~GeV$^2$}}
\put(42,-3){\mbox{$M^2_{R,13} \cdot 10^3$~GeV$^2$}}
\put(84,54){\mbox{\small \bf (b)}}
\put(90,53){\mbox{\small $M^2_{L,23} \cdot 10^3$~GeV$^2$}}
\put(127,-3){\mbox{\small $M^2_{R,23} \cdot 10^3$~ GeV$^2$}}
\end{picture}
\caption{Ranges for parameters inducing lepton number violation.}
\label{fig:parameter}
\end{figure} 

We find values for  $|M^2_{R,ij}|$ up to $8 \cdot 10^3$~GeV$^2$, $|M^2_{L,ij}|$
up to $6 \cdot 10^3$~GeV$^2$ and $|A_{ij} v_d|$ up to 650~GeV$^2$ compatible 
with the 
constraints. In most cases, one of the mass squared parameters is at least
one order of magnitude larger than all the others. However, there is a
sizable part in parameters where at least two of the off-diagonal parameters
have the same order of magnitude as shown in \fig{fig:parameter}.

\section{Signals}

\begin{figure}
\setlength{\unitlength}{1mm}
\begin{picture}(170,63)
\put(0,-3){\mbox{\epsfig{figure=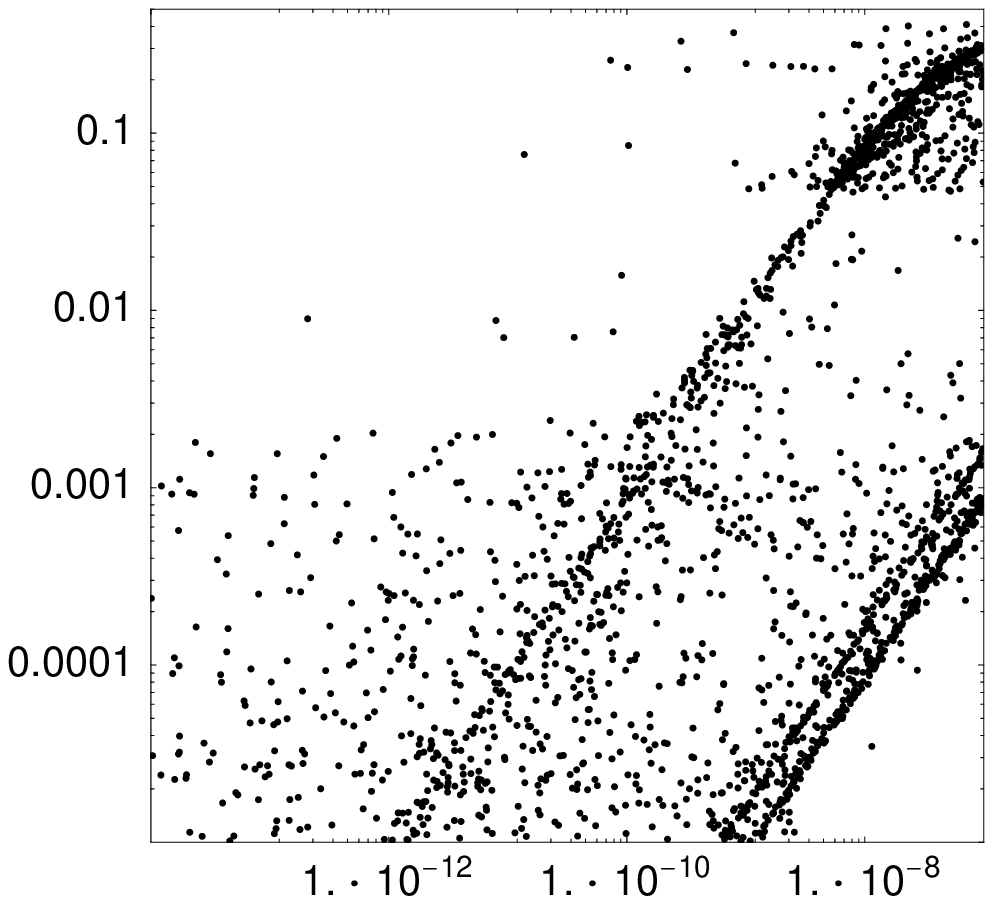,height=6.cm,width=7.5cm}}}
\put(80,-3){\mbox{\epsfig{figure=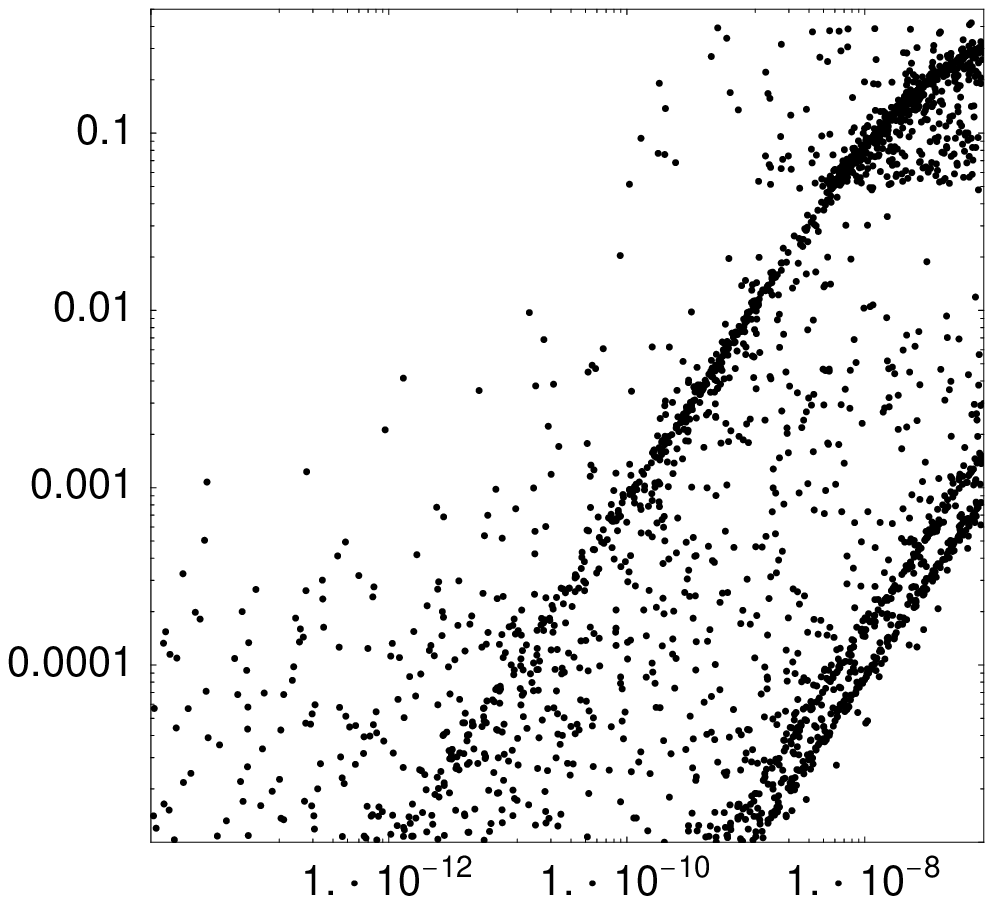,height=6.cm,width=7.5cm}}}
\put(0,58){\mbox{\bf (a)}}
\put(7,57){\mbox{BR$(\tilde \chi^0_2 \to \tilde \chi^0_1 \, e^\pm \tau^\mp)$}}
\put(55,-3){\mbox{BR$(\tau \to e \gamma)$}}
\put(79,58){\mbox{\bf (b)}}
\put(86,57){\mbox{BR
                  $(\tilde \chi^0_2 \to \tilde \chi^0_1 \, \mu^\pm \tau^\mp)$}}
\put(133,-3){\mbox{BR$(\tau \to \mu \gamma)$}}
\end{picture}
\caption{Flavour violating decays of the second lightest neutralino.}
\label{fig:chi2decays}
\end{figure} 

In what follows, we concentrate on possible LFV signals at a 500~GeV
$e^+ e^-$ collider: $e \mu \,  \misset$, 
$e \tau \, \misset$, $\mu \tau \,  \misset$, as well as the 
possibility of additional jets.  We have generated 8000
points consistent with the experimental data, varying
the parameters randomly on a logarithmic scale: $ 10^{-8} \le |A_{ij}|
\le 50$~GeV, $ 10^{-8} \le M^2_{ij} \le 10^4$~GeV$^2$. 
 We consider the following SUSY processes: 
$e^+ e^- \to \tilde l^-_i \tilde l^+_j, \tilde \nu_{i'} \bar{\tilde \nu}_{j'},
\tilde \chi^0_k \tilde \chi^0_m, \tilde \chi^+_n \tilde \chi^-_o$ as
well as stop and Higgs production. We take into account all possible SUSY
 and Higgs cascade decays. We have taken into account ISR- and
SUSY-QCD corrections for the production cross sections.

The main sources for the LFV signal stem from production of sleptons,
sneutrinos and their decays, for example:
\begin{eqnarray}
e^+ e^- \to \tilde l^-_i \tilde l^+_j \to l^-_k l^+_m 2 \tilde \chi^0_1 \, .
\end{eqnarray}
Moreover, also the decays of the second lightest neutralino give an important
contribution as shown in \fig{fig:chi2decays}. There are two main reasons for 
the large
flavour violating branching ratios of $\tilde \chi^0_2$
in some parts of the parameter space:
(i) There is no negative interference terms with a $Z$--boson exchange
as in the case of flavour conserving decays into leptons. (ii) The squarks
are substantially heavier than the sleptons in this scenario.
The cross section for the LFV signal $e \tau \misset$ 
can go up to 250 fb if both 
beams are polarized leading 
to about $10^4$ events with a luminosity of 100~fb$^{-1}$.
 In the case of
two leptons with different flavors and 2 jets we find cross sections up to
1.5~fb \cite{ego}, we have put a veto on
b-jets because of the large background stemming from $t$-quark
production.

\begin{figure}
\setlength{\unitlength}{1mm}
\begin{picture}(170,66)
\put(0,0){\mbox{\epsfig{figure=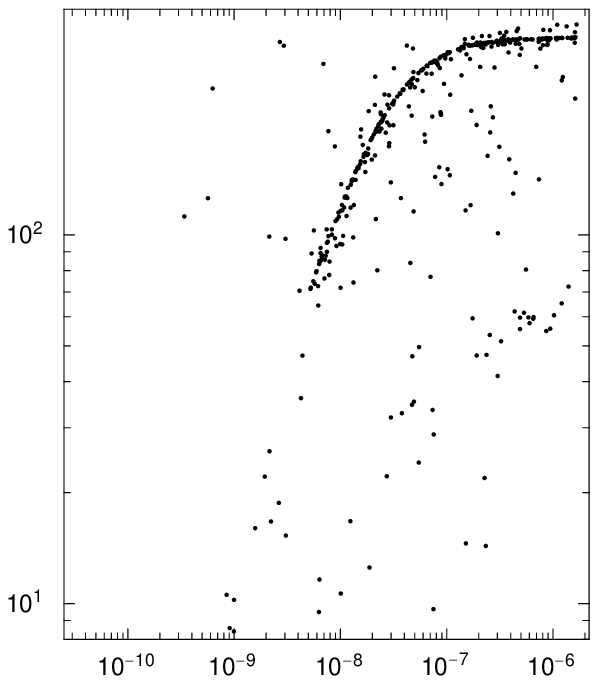,height=5.9cm,width=7.5cm}}}
\put(80,0){\mbox{\epsfig{figure=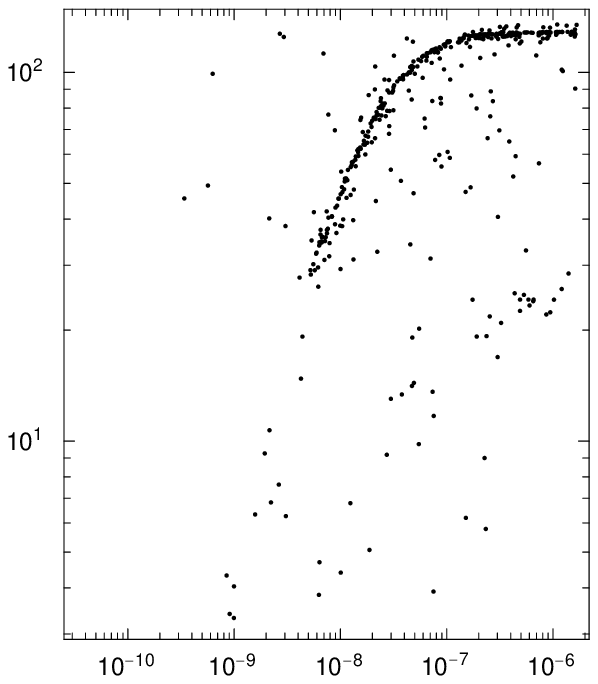,height=5.9cm,width=7.5cm}}}
\put(0,61){\mbox{\bf (a)}}
\put(7,61){\mbox{$\sigma(e^+ e^- \to
e^\pm \tau^\mp \misset)$~[fb]}}
\put(55,-3){\mbox{BR$(\tau \to e \gamma)$}}
\put(79,61){\mbox{\bf (b)}}
\put(87,61){\mbox{$signal/\sqrt{background}$}}
\put(135,-3){\mbox{BR$(\tau \to e \gamma)$}}
\end{picture}
\caption{({\bf a}) Cross section in fb for the signal $e^\pm \tau^\mp \misset$ 
            and  ({\bf b}) the ratio
           signal over square root of background  as a function of 
           BR$(\tau \to e \gamma)$ 
           for $\sqrt{s} = 500$~GeV, $P_{e^-} = 0$ and $P_{e^+} = 0$. In the
           latter case we have assumed an integrated luminosity of 
           100 fb$^{-1}$.}
\label{fig:signal1}
\end{figure}

For the background we take into account all possible SUSY cascade
decays faking the signal and the Standard Model background from
$W$-boson pair production, $t$-quark pair production and $\tau$-lepton
pair production. The SM background has been calculated with the program
Whizard \cite{Kilian:2001qz}.
A SUSY background reaction is, for example, the chain
$\tilde \chi^0_r \to l^-_j \nu_i \tilde \chi^+_s \to l^-_j \nu_i l^+_k
\nu_n \tilde \chi^0_m$. 
In \fig{fig:signal1} we show the cross section of $e^+ e^- \to
e^\pm \tau^\mp \misset$  and 
the corresponding ratio signal over square root of the background
($S/\sqrt{B}$) as a function $BR(\taueg)$ assuming an integrated
luminosity of 100~fb$^{-1}$.  Although no cuts have been applied,
there is in most cases a spectacular signal. The cases
where the ratio $S/\sqrt{B}$ is of order 1 or smaller should
clearly improve, once appropriate cuts are applied. For example, a cut
on the angular distribution of the final state leptons
will strongly reduce the $WW$ background. 
Further cuts as applied in the study of slepton
production \cite{vandervelde} will enhance the ratio  $S/\sqrt{B}$.
The accumulation of points in \fig{fig:signal1} along a band is due to
a large $\tilde e_R$-$\tilde \tau_R$
mixing which is less constraint by $\taueg$ than the corresponding
left-left or left-right mixing.

Let us shortly comment on the situation where neutrino data are not explained
by the see--saw mechanism but due to bilinear terms in the superpotential
breaking R-parity (see e.g.~\cite{numass} and references therein). 
It has been shown
that the additional R--parity breaking contributions to processes such 
as $\mueg$ are negligible \cite{romao}
so that the same ranges of flavour violating parameters as in the study above
are allowed. Thus the same sources for the various signals
plus additional leptons stemming from the LSP decays 
(see e.g.~\cite{neut}) are present.
This clearly will lead to even larger signals, in particular those containing
additional jets. 

\section{Summary}

In conclusion, we have shown that the most general flavour violating structure
of the slepton and sneutrino mass matrix may lead to large lepton flavour
violating signals at a future $e^+ e^-$ collider
-- despite the strong constraints on rare lepton decays.

\section*{Acknowledgements}

This work was supported by the `Fonds zur F\"orderung der
wissenschaftlichen Forschung' of Austria, project No.~P13139-PHY and
the Erwin Schr\"odinger fellowship Nr.~J2095, by the EU TMR Network
Contract No.~HPRN-CT-2000-00149, and partly by the Swiss 'Nationalfonds'.

\end{document}